\documentclass{webofc}
\usepackage[varg]{txfonts}   
\woctitle{?????????}
\begin{document}
\title{Gamma-ray emission from young supernova remnants: hadronic or leptonic?}
%
%

\author{Stefano Gabici\inst{1}\fnsep\thanks{\email{stefano.gabici@apc.univ-paris7.fr}} \and
        Felix Aharonian\inst{2,3}
}

\institute{APC, Univ Paris Diderot, CNRS/IN2P3, CEA/Irfu, Obs de Paris, Sorbonne Paris Cit\'e, France
\and
           Dublin Institute for Advanced Studies, 31 Fitzwilliam Place, Dublin 2, Ireland
\and
           Max-Planck-Institut f\"ur Kernphysik, Postfach 103980, D-69029 Heidelberg, Germany 
          }

\abstract{
The debate on the nature of the gamma-ray emission from young supernova remnants is still open. Ascribing such emission to hadronic rather than leptonic processes would provide an evidence for the acceleration of protons and nuclei, and this fact would fit with the very popular (but not proven) paradigm that supernova remnants are the sources of Galactic cosmic rays. 
Here, we discuss this issue with a particular focus on the best studied gamma-ray-bright supernova remnant: RX~J1713.7-3946. 
}
\maketitle
\section{Introduction}
\label{intro}

The acceleration of protons at supernova remnant (SNR) shocks is accompanied by the emission of gamma-rays, resulting from the decay of neutral pions produced in interactions between the accelerated cosmic rays (CR) and the ambient interstellar matter (see e.g. \cite{pierre} and references therein). It is widely believed that the efficiency of energy conversion into CRs should be at the $\approx 10 \%$ level at SNR shocks, i.e. each SNR should convert $W_{\rm CR} \approx 10^{50}$~erg into CRs. The reason for that is phenomenological: it is the energy needed to sustain the observed intensity of CRs against their escape from the Galaxy \cite{gaisser}. 
An estimate of the very high energy (VHE) gamma-ray emission resulting from the interactions between the CRs accelerated in a SNR and the ambient gas of density $4 \times n_{gas}$ (the factor of 4 accounts for the compression of the gas at the SNR shock) is \footnote{For convenience, all the estimates done in this section relies on the assumption of a $E^{-2}$ spectrum for both protons and electrons. It is trivial to consider the more general case $E^{-s}$ ($s \gtrsim 2$ seems to better agree with observations of SNRs and of CRs) and check that the main claims made here remain substantially correct.}:
\begin{equation}
\label{eq:pp}
E^2_{\gamma} F_{\gamma} \approx \frac{W_{\rm CR}}{\ln(E_{\rm max}/E_{\rm min}) ~ \tau_{\pi^0} ~ (4 \pi d^2)} =  4 \times 10^{-11} \left( \frac{W_{\rm CR}}{10^{50}~{\rm erg}} \right) \left( \frac{n_{gas}}{\rm cm^{-3}} \right) \left( \frac{d}{\rm kpc} \right)^{-2} {\rm erg/cm^2/s}
\end{equation}
where $\tau_{\pi^0} \approx 1.6 \times 10^8 (n_{gas}/{\rm cm}^3)$~yr is the energy loss time due to neutral pion production and $d$ the distance to the SNR. In computing Eq.~\ref{eq:pp} we implicitly assumed that the CR spectrum is $\propto E^{-2}$ and extends from $E_{min} = 1$~GeV to $E_{max} = 4$~PeV, which is the energy of the CR knee. This is a large flux, well within the reach of current Cherenkov telescopes, unless the SNR is too far away or the ambient density is much lower than the average one in the Galaxy \cite{dav}.  
A spectrum much steeper than $E^{-2}$ and/or a cutoff in the spectrum at an energy much lower than that of the knee would result is a dramatic suppression of the VHE gamma-ray flux, but these spectral feature would be in contradiction with the SNR paradigm for the origin of CRs.
For this reason, the detection of SNRs in VHE gamma-rays was long awaited and considered a test for CR origin \cite{dav}. However, the detection of several SNRs in VHE gamma-rays cannot be considered a proof of CR acceleration because leptonic processes (most notably inverse Compton scattering) could as well explain the observed emission \cite{felixSNRs}.

The contribution to the gamma-ray flux from inverse Compton scattering of electrons in the cosmic microwave background radiation can be computed by recalling that photons of energy of $\approx 1$~TeV are produced by electrons of energy $E_e \approx 20$~TeV. At this energy the rate of production of TeV photons is $\tau_{\rm IC}^{-1} \approx (5 \times 10^4 {\rm yr})^{-1}$. If the spectrum of electrons accelerated at a SNR shock is $K_{ep}$ times that of protons, then the contribution to the $\approx 1$~TeV gamma-ray flux from inverse Compton scattering is $\approx K_{ep} W_{\rm CR}/2 \tau_{\rm IC} \ln(E_{max}/E_{min}) (4 \pi d^2)\approx 2 \times 10^{-11} (W_{CR}/10^{50}~{\rm erg}) (d/{\rm kpc})^{-2} (K_{ep}/10^{-3})$~erg/cm$^2$/s. From which the ratio of leptonic to hadronic contribution to the VHE gamma-ray emission at 1 TeV can be deduced:
\begin{equation}
\label{eq:epratio}
\phi_{\rm IC/pp}(1~{\rm TeV}) \approx 0.5 ~\left( \frac{n_{gas}}{\rm cm^{-3}} \right)^{-1} \left( \frac{K_{ep}}{10^{-3}} \right)
\end{equation}
which indicates that in normal interstellar medium conditions the leptonic contribution dominates unless electrons are less than $\approx 0.1 \%$ abundant than protons. In other words, inverse Compton scattering is much more efficient than proton-proton interactions in producing gamma rays and thus the leptonic emission from a SNR might well exceed the hadronic one even in the presence of a significant acceleration of protons. Hence, the difficulty of ascribing unambiguously the VHE gamma-ray emission from SNRs to hadronic processes.

The estimate given above is valid under the assumption that the shape of electron and proton spectra are identical. This is true only if the electron spectrum is not affected by energy loss processes\footnote{Energy losses of protons have characteristic times well in excess than the SNR age and can be safely neglected.}. In SNRs the main cooling channel for electrons is synchrotron radiation, which proceeds at a rate $\tau_{syn} \approx 6 \times10^3 (B/10~\mu{\rm G})^{-2} (E_e/20~{\rm TeV})^{-1}$~yr. This implies that in young SNRs of age $\tau_{age}$ of the order of few thousand years the multi-TeV spectrum of electrons is affected by synchrotron losses if the magnetic field is $\gtrsim 10~\mu{\rm G}$.
In this case, the ratio between leptonic and hadronic emission is significantly reduced with respect to the value computed in Eq.~\ref{eq:epratio}, because a break forms in the electron spectrum at an energy $E^*_e$ that satisfies the condition $\tau_{age} \approx \tau_{syn}(E_e^*)$.

The spectrum of electrons accelerated at a SNR shock can thus be described as follows: at low energies, where both electrons and protons are unaffected by energy losses, the electron spectrum is identical in shape to the proton one, but is rescaled by the factor $K_{ep}$ which accounts for the different acceleration efficiency of the two species. If the magnetic field strength is large enough, a break may appear in the electron spectrum at an energy $E_e^* \approx 1 ~ (B/100~\mu{\rm G})^{-2} (\tau_{age}/{\rm kyr})^{-1}$~TeV. In gamma rays, the break in the inverse Compton spectrum would appear at a photon energy $E_{\gamma}^* \approx 4 ~ (B/100~\mu{\rm G})^{-4} (\tau_{age}/{\rm kyr})^{-2}$~GeV. At energies larger than the break, the electron and gamma-ray spectra steepen by one power and half power of energy, respectively, and proceed until a maximum energy which can be computed by equating the energy loss time (determined by synchrotron radiation) to the acceleration time at the shock $D_B(E_e)/u_s^2$, where $D_B = (1/3) R_L c$ is the Bohm diffusion coefficient, $R_L$ the particle Larmor radius, $c$ the speed of light, and $u_s$ the shock speed. This gives $E_e^{max} \approx 30~(B/100~\mu {\rm G})^{-1/2} (u_s/1000~{\rm km/s})$~TeV, which corresponds to a gamma-ray photon energy of a few TeV in the inverse Compton spectrum. Thus, a cutoff in the electron spectrum at energies $< 30$~TeV would prevent the production of VHE gamma rays through inverse Compton scattering, unless soft radiation fields other than the cosmic microwave radiation (e.g. infrared, optical ... ) are present.

Having identified all the main physical quantities that shape the hadronic and leptonic contribution to the gamma-ray spectrum of SNRs, we can proceed discussing a specific object, RX~J1713.7-3946, in an attempt to understand the origin of its gamma-ray emission.

\section{The SNR RX~J1713.7-3946}

RX~J1713.7-3946 is the best studied SNR in VHE gamma-rays and the hadronic or leptonic nature of its emission is still debated (for a non-exhaustive list of references see \cite{berezhko,giovanni,volodia,don,finke,meRXJ}).
In the following we discuss this issue by listing the various arguments that have been brought forward in favor or a hadronic or leptonic origin of the emission.

\subsection{Arguments in favor of an hadronic gamma-ray emission}

\subsubsection{Evidence for magnetic field amplification}

According to shock acceleration theory, the acceleration of CRs at SNR shocks must be accompanied by the amplification of the magnetic field up to values of hundreds of milliGauss or even milliGauss. In order to have an effective amplification of the field, an efficient acceleration of CR hadrons is required \cite{bell2004,bell2013}. 
This finding received strong observational support with the detection of thin X-ray synchrotron filaments from several SNR shocks. In an amplified field, electrons accelerated at shocks are cooled much more rapidly than the rate at which they are advected downstream of the shock. Thus, they radiate synchrotron X-rays in the close proximity of the shock and their emission appears as a very thin filament. 
The thinness of filaments allows to constrain the magnetic field strengths in the hundreds of microGauss range from a number of young SNRs (see \cite{jacco} for a review). Though the presence of X-ray filaments in RX~J1713.7-3946 was revealed by Chandra observations \cite{yas2}, the most spectacular evidence for magnetic field amplification in this object came from the observation, also made by Chandra, of fast (year-scale) variability of the synchrotron emission of some small (sub-parsec scale) X-ray knots \cite{yas2}.
If the variability is interpreted as the result of the fast cooling of freshly accelerated electrons, a magnetic field of $\approx 1$~mG is inferred \cite{yas2}. Alternatively, the rapid variability of the X-ray emission can be explained by a fluctuating magnetic field. This has been indeed proposed and discussed in \cite{bykov}, where the authors, in order to explain the variability, adopted a $100~\mu$G field, which is lower than that inferred in \cite{yas2} but still significantly amplified.

It has to be stressed that the evidence for an amplified magnetic field is limited to a very small fraction of the SNR volume (filaments and knots) and thus it is difficult to reach strong conclusions on a global level. For example, it is unclear if CR protons are accelerated with large efficiencies in the whole shock surface or limited to a small fraction of it. Also, very little constrains are available for the strength of the magnetic field in the whole remnant. If the volume averaged value of the magnetic field is significantly larger $\approx 10~ \mu$G then a leptonic interpretation of the gamma-ray emission from RX~J1713.7-3946 is problematic, because in this case electrons would radiate mainly through synchrotron radiation rather than inverse Compton scattering. On the other hand, if the magnetic field outside filaments and knots is at the level of $\approx 10 ~\mu$G then the leptonic interpretation would be definitely favored.

To conclude, high resolution X-ray observations of SNRs, and in particular of RX~J1713.7-3946, are generally interpreted as a very strong evidence for an efficient acceleration of CR hadrons from at least a part of SNR shocks. It is tempting to link such an efficient acceleration efficiency with a correspondingly high level of the hadronic gamma-rays from the remnants. However, the ambient gas density also plays a crucial role here (see Eq.~\ref{eq:pp}), and it can be easily shown that if the target density is as low as 0.1 cm$^{-3}$ even an effective conversion of energy into CRs ($\approx 3 \times 10^{50}$~erg in form of CR protons) would not suffice to provide a significant hadronic contribution to the observed gamma-ray flux \cite{fermi}.

\subsubsection{Spatial correlation of gamma-ray emission and gas density}

The search for spatial correlations between gamma-ray brightness and gas column density is a powerful tool that can serve to discriminate amongst hadronic and leptonic emission. This is because a correlation with the ambient gas density is naturally expected in a hadronic scenario where CR protons produce gamma rays in interactions with the ambient gas. Such a correlation is not granted (but at the same time not excluded!) in leptonic models. 

A spatial correlation of SNRs and dense gas is likely if the SNR progenitor is a supernova of type II, which is expected to explode inside a molecular cloud as a result of the very fast evolution of a very massive star \cite{thierry}. Indeed, a type II progenitor for RX~J1713.7-3946 seems to be suggested by observations \cite{slane}, and a correlation with dense molecular gas has been claimed \cite{yoshi}. The significance of the correlation has been found to increase as a result of detailed studies that considered the total gas density (atomic plus molecular) in the region \cite{fukui}. Also, the ambient gas within the SNR shell appears to be structured in small (sub-parsec) and dense (up to at least $10^4$~cm$^{-3}$) clumps \cite{nigel}. The presence of a significant amount of gas in the SNR shell might of course provide a target for CR proton-proton interactions and enhance the hadronic contribution to the gamma-ray emission \cite{fukui}.

\subsection{Arguments in favor of a leptonic gamma-ray emission}

\subsubsection{Spectral shape of the gamma-ray emission}

The spectral shape of the gamma-ray emission from RX~J1713.7-3946 is probably the most popular argument presented in favor of a leptonic origin of the emission. As described in the Introduction, if the magnetic field is amplified, two features are expected in the leptonic gamma-ray spectrum of SNRs: {\it i)} a break at energy $E_e^*$ due to the fact that particles above a given energy cool via synchrotron emission on a time scale shorter than the age of the remnant, and {\it ii)} a cutoff at an energy $E_e^{max}$ which represents the maximum energy a particle can be accelerated to before losing energy via synchrotron radiation. If we consider an injection spectrum of electrons proportional to $E_e^{-2}$, then the electron (gamma-ray) spectrum is $\propto E_e^{-2}$ ($\propto E_{\gamma}^{-1.5}$) below the break and $\propto E_e^{-3}$ ($\propto E_{\gamma}^{-2}$) above it.

The picture described above changes if the magnetic field at the SNR is weak. This is because both the break and the cutoff energies increase for weaker magnetic fields, but in two different ways, i.e., $E_e^* \propto B^{-2}$ and $E_e^{max} \propto B^{-1/2}$, implying that a magnetic field strength exists below which the break in the spectrum disappears. Such a field is $\approx 10~\mu$G (see Introduction), and for field strength smaller or equal to this value the electron (gamma-ray) spectrum becomes a simple power law in energy with index -2 (-1.5). 

Indeed, a weak magnetic field is required for the leptonic interpretation to be viable. Synchrotron X-ray emission has been detected from RX~J1713.7-3946 at a flux level of $E_X^2 F_X \approx 10^{-10}$~erg/cm$^2$/s, while the VHE gamma-ray emission $E_{\gamma}^2 F_{\gamma}$ lays roughly one order of magnitude below that. 
If both X-rays and gamma rays are produced by the same electrons through synchrotron and inverse Compton scattering in the cosmic microwave background, respectively, then their flux ratio should reflect the ratio between the energy density in magnetic field $\omega_B = B^2/8 \pi \approx 2.5~(B/10~\mu {\rm G})^2$~eV/cm$^3$ and in the cosmic microwave background $\omega_{CMB} \approx 0.25$~eV/cm$^3$:
\begin{equation}
\frac{F_X}{F_{\gamma}} \approx 10 \left( \frac{B}{10 ~ \mu {\rm G}} \right)^2
\end{equation}
which implies that a field strength of $\approx 10~ \mu$G is required to fit data.

\begin{figure}
\centering
\sidecaption
\includegraphics[width=6cm,clip]{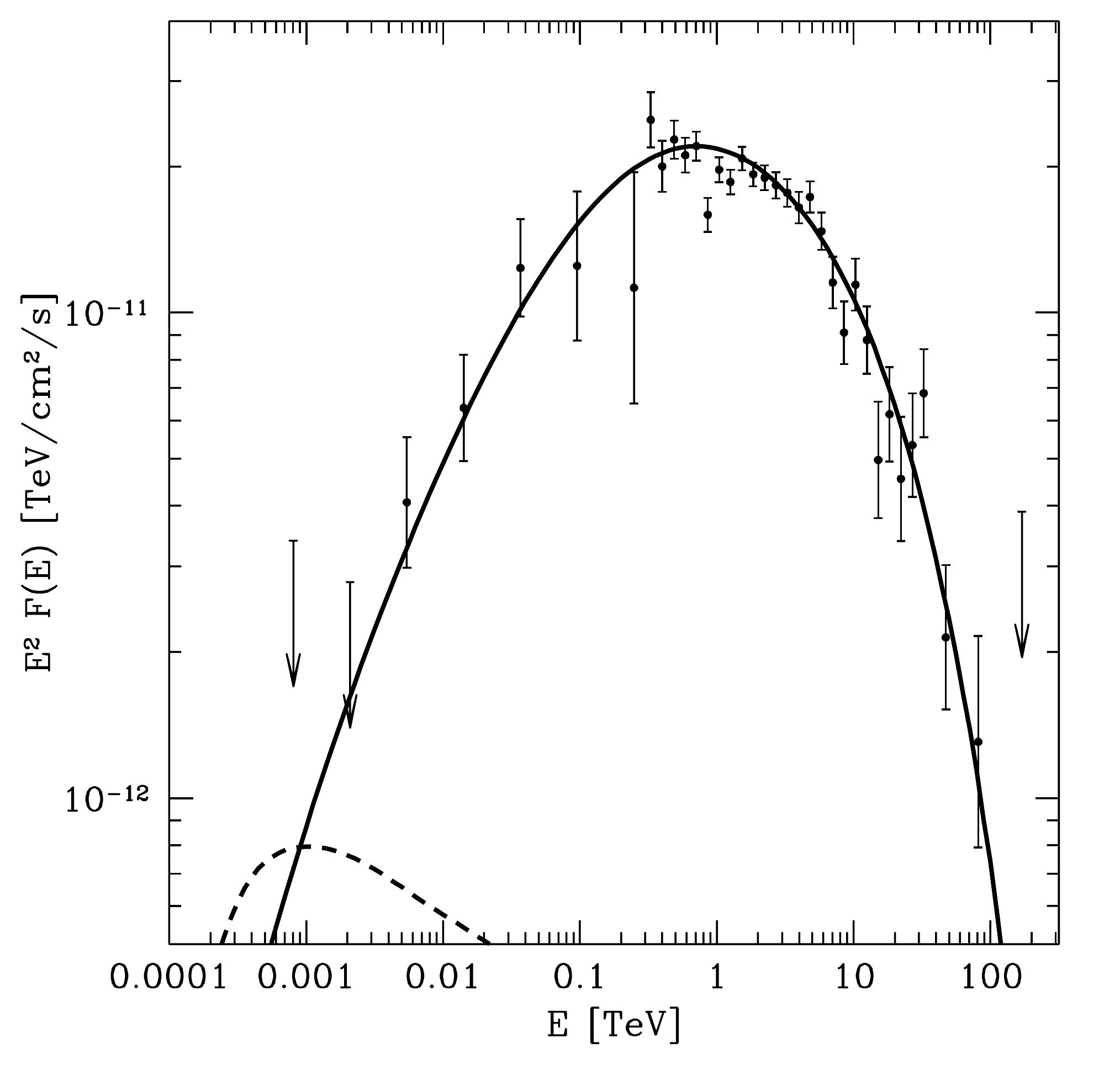}
\caption{Hadronic model for RX~J1713.7-3946, under the assumption that the remnant originated from a type II supernova explosion in a molecular cloud \cite{meRXJ}. The SNR is assumed to expand in a rarefied ($\sim 2 \times 10^{-2}$~cm$^{-3}$) cavity inflated by the wind of the progenitor star. The wind swept away all the matter but the densest parts ($\gtrsim 10^3$~cm$^{-3}$), concentrated in small ($\sim 0.1$~pc) clumps. Such dense clumps survive intact the passage of the SNR shock. CRs accelerated at the SNR shock diffusively penetrate into the clumps. Bohm diffusion in a $\approx 100 ~ \mu$G field is assumed to operate at the interface between the clumps and the diffuse medium in the SNR. The solid line represents the predicted emission from the clumps, while the dashed line that from the diffuse gas inside the SNR. Here, an explosion energy of $10^{51}$ ergs, a CR acceleration efficiency of $10\%$ and a spectrum of accelerated CRs at the shock $\propto E^{-2.2}$ have been assumed. To match the data, the total mass in the clumps has to be $\approx 550~M_{\odot}$. Data points are from FERMI \cite{fermi} and HESS \cite{hess}.}
\label{fig}       
\end{figure}

To summarize, a simple, one-zone model leptonic interpretation of the spectrum of RX~J1713.7-3946 requires a weak magnetic field, and a rising spectral energy distribution of gamma-ray photons below the cutoff: $E_{\gamma}^2 F_{\gamma} \propto E_{\gamma}^{0.5}$. On the other hand, if the magnetic field is $\gg 10~ \mu$G, an hadronic interpretation seems favored, and for a typical $\propto E^{-2}$ injection spectrum of protons one would expect a flat spectral energy distribution of gamma-rays from neutral pion decay: $E_{\gamma}^2 F_{\gamma} \propto E_{\gamma}^0$. This is the reason why the detection by {\it Fermi} of a rising spectral energy distribution in the GeV domain from RX~J1713.7-3946 (see \cite{fermi} and Fig.~\ref{fig}) has been interpreted as a strong evidence for the fact that the gamma-ray emission is indeed leptonic. Following the same reasoning, the much steeper gamma-ray spectrum detected from the Tycho SNR has been interpreted as an evidence for hadronic emission \cite{fermitycho,giovannitycho}.

In fact, it has been soon realized that a one-zone leptonic model fails to obtain a good fit to the gamma-ray data for RX~J1713.7-3946. The reason is that the expected spectrum has a peak much narrower than the observed one (see e.g. Fig.~3 in \cite{felixSNRs}). This problem can be solved either by assuming that two populations of relativistic electrons exist (e.g. accelerated at the forward and reverse shock, respectively?) \cite{finke,tanaka}, or by invoking the presence of a strong infrared radiation field in the SNR that would provide a target for inverse Compton interactions distinct than the cosmic microwave background \cite{porter}. However, such a strong additional radiation field seems quite difficult to be justified (see e.g. \cite{giovanni}).

\subsubsection{Absence of thermal X-ray emission}

The absence of thermal Bremsstrahlung continuum emission in the X-ray spectrum of RX~J1713.7-3946 was claimed to be an evidence against the hadronic origin of the gamma rays \cite{katz}. In order to have effective CR hadronic interactions the ambient gas density must not be too low, since the hadronic emission scales as the target gas density. On the other hand, if this is the case one should also expect a powerful X-ray thermal emission from the gas heated by the SNR shock (thermal emission scales as the gas density squared). 
In fact, it has been shown that the problem comes mainly from the non detection of X-ray thermal lines, rather than continuum emission. Lines should overcome the observed synchrotron emission of RX~J1713.7-3946 unless the ambient density is very low, $\lesssim 0.05$~cm$^{-3}$. Since no thermal emission is observed (neither continuum nor lines) hadronic models are challenged \cite{don}.

The hadronic interpretation would still be viable if the gas is not heated to the high temperatures required for the production of X-rays. This could happen if the shock converts a large fraction (significantly above $\approx 10\%$) of the incoming kinetic energy into CRs, since in this case not enough energy would be left to heat the gas (e.g. \cite{donheating,blasi,luke,hevelin}).
However, recent simulations of the injection phase of particles at shocks suggest for the CR acceleration efficiency a value of $\approx 10\%$ \cite{damiano}, implying that significan gas heating must occur. Another interesting possibility to avoid gas heating involves the presence of dense clumps in the medium surrounding the SNR and will be discussed below.



\subsection{A hadronic model for RX~J1713.7-3946}

It has been argued in \cite{volodia} and \cite{inoue} that the origin of the gamma-ray emission from RX~J1713.7-3946 may well be hadronic if the SNR expands in a clumpy medium. The simulations performed in \cite{inoue} suggests that this is a quite natural situation if the SNR progenitor is a type II supernova exploding in a molecular cloud, as RX~J1713.7-3946 is believed to be \cite{slane}. The powerful wind from the progenitor star would sweep away the dense gas creating a cavity  characterized by a gas density of the order of $\approx 10^{-2}$~cm$^{-3}$. However, the densest cores (of density above $\approx 10^3$~cm$^{-3}$) in the molecular cloud would survive into the cavity. Remarkably, such clumps would also survive the SNR shock passage. This happens because a SNR shock of velocity $u_s$ that hits a dense clump generates a shock into the clump itself, with a velocity $u_c \approx u_s/a$, where $a \approx (10^3 {\rm cm}^{-3}/10^{-2} {\rm cm}^{-3})^{1/2} \approx 300$ is the square root of the ratio between the clump to inter clump density \cite{klein}. Thus, the SNR shock is virtually stalled at the clump border and the clump remains {\it unshocked}. 

Remarkably, this fact solves one of the problems of hadronic models. The absence of thermal emission can be explained if the gas mass within the SNR is dominated by the contribution of dense clumps. Given that clumps remain unshocked and thus cold no appreciable X-ray thermal emission is expected.

The presence of clumps can also solve the problem related to the observed gamma-ray spectrum of RX~J1713.7-3946, which is generally considered too hard to be explained by hadronic interactions. According to simulations, the interaction between the SNR and the clumps would lead to a strong amplification of the turbulent magnetic field at the interface between the clumps and the diffuse gas \cite{inoue}. Such a turbulent magnetic field would prevent low energy CRs (accelerated at the SNR shock and confined within the SNR shell) to penetrate the clumps over a time scale shorter than the SNR age ($\approx 1600$~yr). On the other hand, high energy CRs would penetrate the clumps, their transport being characterized by a faster diffusion coefficient. The outcome of this is that the CR spectrum inside clumps is expected to be much harder than that of CRs in the SNR shell. If the total mass of the gas in the shell is dominated by the clumps, then the gamma-ray spectrum observed from the SNR would be hard, as the underlying CR spectrum inside clumps. 

Following these arguments, a detailed model for RX~J1713.7-3946 has been developed in \cite{meRXJ}, demonstrating that an excellent fit to the data can be obtained with a hadronic model (see Fig.~\ref{fig}).

\section{Conclusions}

The origin of the gamma-ray emission from RX~J1713.7-3946 is not well established yet. A leptonic scenario requires a low magnetic field strength ($\approx 10 ~ \mu$G), and either two populations of relativistic electrons or two populations of target soft photons for inverse Compton interactions. On the other hand, a hadronic scenario implies a much stronger magnetic field and requires the presence of a structured ambient medium with dense clumps embedded in a tenuous diffuse gas. The detection or non detection of neutrinos will allow to discriminate amongst these two scenarios. For the hadronic scenario the detection of neutrino is well within the reach of a km$^3$-scale detector located in the northern hemisphere \cite{francesco}. 

Also the detection of radiation from beyond the SNR shell would provide strong support to the hadronic scenario. Such a radiation is expected from the vicinity of the shell as the result of the interactions of escaping CR protons in the ambient gas \cite{atoyan,mecloud}. For RX~J1713.7-3946 this would result in the presence of a  faint diffuse gamma-ray emission surrounding the SNR, of brightness at the limits of detectability of current Cherenkov instruments, but within the reach of the future Cherenkov Telescope Array \cite{sabrinona}. 
Remarkably, also X-ray telescopes like NuSTAR and Astro-H can contribute to the search of escaping CR protons by detecting the synchrotron emission produced by the secondary electrons produced by runaway CRs in interactions with the ambient gas \cite{astroH}.

\textit{Acknowledgements:} SG acknowledges Paolo Piattelli and all the organizers of the RICAP conference for the invitation and the kind hospitality. He also acknowledges support under the LabEx UnivEarthS program at Sorbonne Paris Cit\'e (ANR-10-LABX-0023/ANR-11-IDEX-0005-02) and the {\it Action F\'ed\'eratrice CTA} at the Observatory of Paris.

%
%
%

\end{document}